\documentstyle[12pt,psfig]{article}
\newcommand{\HI}{H\thinspace\protect\footnotesize I\protect\normalsize} 

\newcommand{\hi}{H\thinspace{\protect\scriptsize I}}

\newcommand{\etal}{{et\,al.}}    % not in italics for PASA
\newcommand{\cf}{{cf.}}          % not in italics for PASA
\newcommand{\eg}{{e.g.}}         % not in italics for PASA
\newcommand{\ie}{{i.e.}}         % not in italics for PASA
\newcommand{\msun}{\mbox{${\rm M}_\odot$}}

\newcommand{\ga}{\,\raisebox{-0.4ex}{$\stackrel{>}{\scriptstyle\sim}$}\,}
\newcommand{\la}{\,\raisebox{-0.4ex}{$\stackrel{<}{\scriptstyle\sim}$}\,}

\def\deg{{^\circ}}
\def\arcmin{\hbox{$^\prime$}}
\def\arcsec{\hbox{$^{\prime\prime}$}}

\def\fm{\hbox{$.\!\!^{\rm m}$}}

\def\fdg{\hbox{$.\!\!^\circ$}}
\def\farcm{\hbox{$.\mkern-4mu^\prime$}}

\def\micron{\hbox{$\mu$m}}
\def\reference{\parskip 0pt\par\noindent\hangindent 0.5 truecm}
\def\kms{km ${\rm s}^{-1}$}
\begin{document}
\title{Mapping the Hidden Universe:\\ 
The Galaxy Distribution in the Zone of Avoidance}

\author{Ren\'ee C. Kraan-Korteweg $^1$ \and
Sebastian Juraszek $^{2,3}$
}

\date{}
\maketitle

{\center
$^1$ Departamento de Astronom\1a, Universidad de Guanajuato, Apartado
Postal 144, Guanajuato GTO 36000, Mexico \\
kraan@astro.ugto.mx\\[3mm]
$^2$ School of Physics, University of Sydney, NSW 2006 Australia \\[3mm]
$^3$ ATNF, CSIRO, PO Box 76, Epping, NSW 2121, Australia \\
sjurasze@atnf.csiro.au\\[3mm]

\begin{abstract} 
Due to the foreground extinction of the Milky Way, galaxies become
increasingly faint as they approach the Galactic Equator creating a
``zone of avoidance'' (ZOA) in the distribution of optically visible
galaxies of about 25\%. A ``whole-sky'' map of galaxies is essential,
however, for understanding the dynamics in our local Universe, in
particular the peculiar velocity of the Local Group with respect to
the Cosmic Microwave Background and velocity flow fields such as in
the Great Attractor (GA) region. The current status of deep optical galaxy
searches behind the Milky Way and their completeness as a function of
foreground extinction will be reviewed. It has been shown that these
surveys -- which in the mean time cover the whole ZOA (Fig.~\ref{cor}) --
result in a considerable reduction of the ZOA from extinction levels
of A$_{\rm B} = 1\fm0$ (Fig.~\ref{ait}) to A$_{\rm B} = 3\fm0$
(Fig.~\ref{aitc}). In the remaining, optically opaque ZOA, systematic
\hi\ surveys are powerful in uncovering galaxies, as is demonstrated
for the GA region with data from the full sensitivity Parkes Multibeam
\hi\ survey ($300\deg \le \ell \le 332\deg$, $|b| \le 5\fdg5$,
Fig.~\ref{MBGA}).
\end{abstract}

}
\bigskip

\section{Introduction}
In 1988, Lynden-Bell \& Lahav for the first time prepared a
``whole-sky'' distribution of the extragalactic light to map the
density enhancements in the local Universe, to compare them to cosmic
flow fields and to determine the gravity field on the Local
Group. Assuming that light traces mass, this could be realized through
a diameter-coded distribution of galaxies larger than ${\rm D} \ga
1\farcm0$ taken from the following galaxy catalogs: the Uppsala
General Catalog UGC (Nilson 1973) for the north ($\delta \ge
-2\fdg5$), the ESO Uppsala Catalog (Lauberts 1982) for the south
($\delta \le -17\fdg5$), while the missing strip ($-17\fdg5 < \delta <
-2\fdg5$) was filled with data from the Morphological Catalog of
Galaxies MCG (Vorontsov-Velyaminov \& Archipova 1963-74).

\begin{figure}
\begin{center}
\hfil \psfig{file=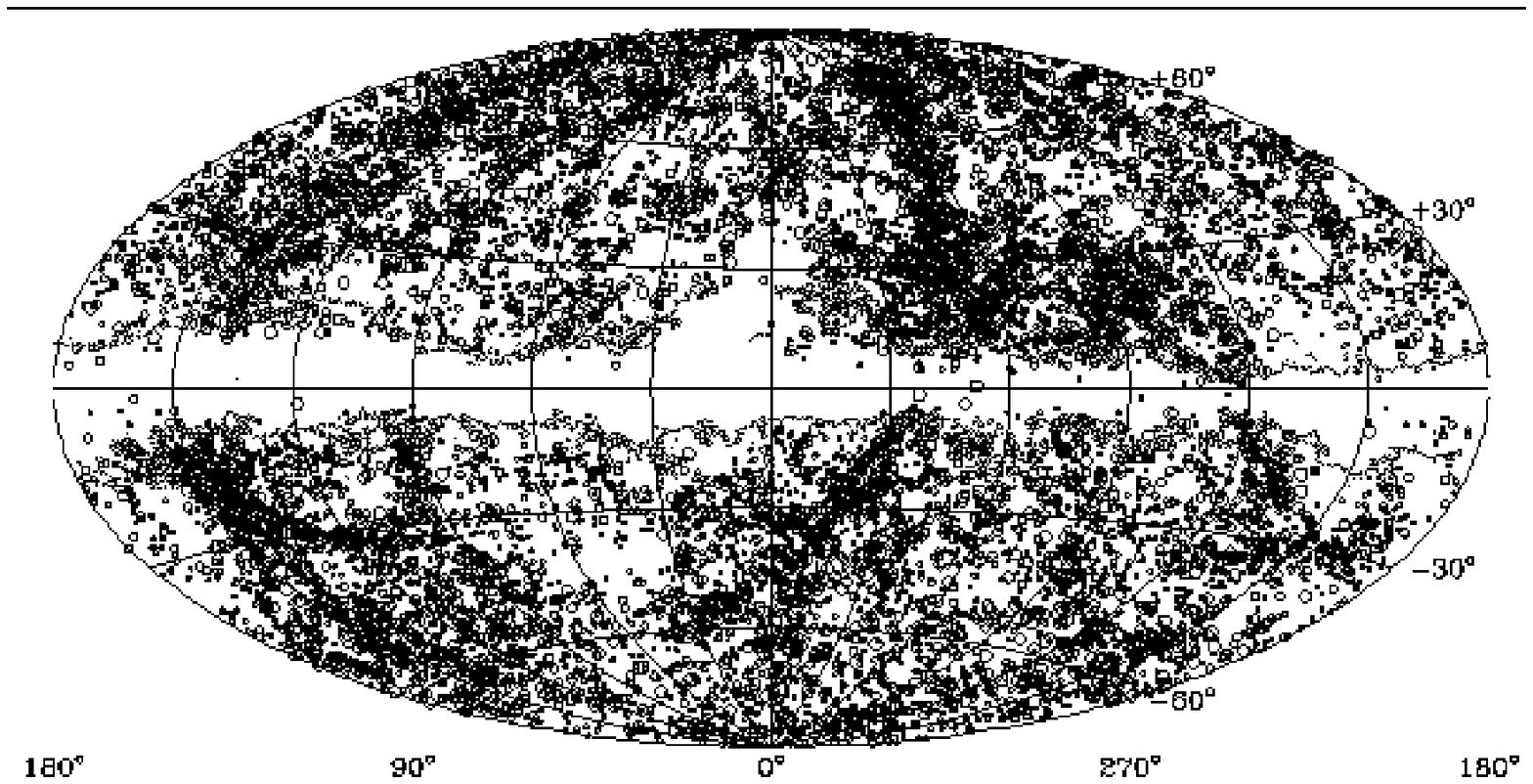,width=16cm} \hfil
\caption
{Aitoff equal-area projection in Galactic coordinates of galaxies with
${\rm D}\ge1\farcm$3. The galaxies are diameter-coded: small circles
represent galaxies with $1{\farcm}3 \le {\rm D} < 2\arcmin$, larger
circles $2\arcmin \le {\rm D} < 3\arcmin$, and big circles ${\rm D}
\ge 3\arcmin$. The contour marks absorption in the blue of ${\rm A_B}
= 1\fm0$ as determined from the Schlegel \etal\ (1998) 
dust extinction
maps.  The displayed contour surrounds the area where the galaxy
distribution becomes incomplete (the ZOA) remarkably well.}
\label{ait}
\end{center}
\end{figure}

Because these optical galaxy catalogs are limited to the largest
galaxies they become increasingly incomplete close to the Galactic
equator where the dust thickens, reducing the apparent diameters
of the galaxies. Added to this are the growing number of
foreground stars which fully or partially block the view of galaxy
images.  This is clearly demonstrated in Fig.~\ref{ait}, where such a
light distribution is presented in an Aitoff equal-area projection
centered on the Galactic plane. The same corrections as advocated in
Lahav (1987) have been applied to homogenize the data of the three
different galaxy catalogs, \ie\ ${\rm D_{25} = 1.15 \cdot D_{UGC},
D_{25} = 0.96 \cdot D_{ESO}}$ and ${\rm D_{25} = 1.29 \cdot D_{MCG}}$.
A cut-off at ${\rm D} =1\farcm3$ was imposed -- the completeness limit
of the respective catalogs according to Hudson \& Lynden-Bell (1991).

Fig.~\ref{ait} clearly displays the irregularity in the distribution
of galaxies in the nearby Universe with its dynamically important
density enhancements such as the Local Supercluster visible as a
circle (the Supergalactic Plane) centered on the Virgo cluster at
$\ell=284\deg, b=74\deg$, the Perseus-Pisces chain bending into the
ZOA at $\ell=95\deg$ and $\ell=165\deg$, the general overdensity in
the Cosmic Microwave Background dipole direction
($\ell=280\deg,b=27\deg$, Kogut \etal\ 1993) and the Great Attractor
region centered on $\ell=320\deg, b=0\deg$ (Kolatt, Dekel \& Lahav 1995) with
the Hydra ($270\deg,27\deg$), Antlia ($273\deg,19\deg$), Centaurus
($302\deg,22\deg$) and Pavo ($332\deg,-24\deg$) clusters. 

Most conspicuous in this distribution is, however, the very broad,
nearly empty band of about 20$\deg$: the Zone of Avoidance. Comparing
this band with the 100\micron\ dust extinction maps of the DIRBE
experiment (Schlegel, Finkbeiner \& Davis 1998) we found that the ZOA -- the area
where our galaxy counts are severely incomplete -- is described almost
perfectly by the extinction contour ${\rm A_B} = 1\fm0$ (where ${\rm
A_B} = 4.14 \cdot {\rm E(B-V)}$, Cardelli, Clayton \& Mathis 1989).

\section{Deep Optical Galaxy Searches}
Systematic deeper searches for ``partially obscured'' galaxies -- down
to fainter magnitudes and smaller dimensions (${\rm D} \ga 0\farcm1$)
than existing catalogs -- have been performed with the aim to reduce
this ZOA. Using existing sky surveys (POSS I and POSS II in the north
and the ESO/SERC surveys in the south), the whole ZOA has in the mean
time been visually surveyed for galaxies. Here, examination by eye is
still the best technique. A separation of galaxy and star images can
as yet not be done by automated measuring machines (\eg\ COSMOS or APM)
on a viable basis below $|b| \la 10\deg-15\deg$ though surveys by eye
are clearly both very trying and time consuming, and maybe not as
objective.

\begin{figure}
\begin{center}
\hfil \psfig{file=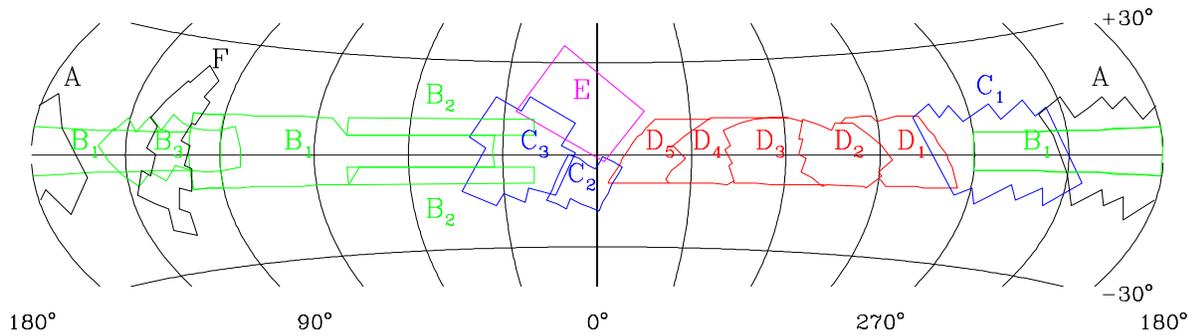,width=16cm} \hfil
\caption
{An overview of the different optical galaxy surveys in the ZOA centered
on the Galaxy. The labels identifying the search areas are explained 
in the text. Note that the surveyed regions  entirely cover the ZOA
defined by the foreground extinction level of ${\rm A_B} = 1\fm0$ in
Fig.~\ref{ait}.}
\label{cor}
\end{center}
\end{figure}

The various surveyed regions are displayed in Fig.~\ref{cor}. Details 
and results on the uncovered galaxy distributions for the various
regions are discussed in:

A:  the Perseus-Pisces Supercluster by Pantoja 1997;\\
B$_{1-3}$: the northern Milky Way (B$_1$ by Seeberger \etal\ 1994,
Seeberger, Saurer \& Weinberger 1996, Lercher, Kerber \& Weinberger 1996, 
Saurer, Seeberger \& Weinberger 1997, and
Seeberger \& Saurer 1998 from POSS I; B$_2$ by Marchiotti, Wildauer \&
Weinberger 1999 from POSS
II; B$_3$ by Weinberger, Gajdosik \& Zanin 1999 from POSS II);\\
C$_{1-3}$: the Puppis region  by Saito \etal\ 1990, 1991 [C$_1$], 
the Sagittarius/Galactic region by Roman, Nakanishi \& Saito 1998 [C$_2$],
and the Aquila and Sagittarius region by Roman \etal\ 1996 [C$_{3}$]; \\ 
D$_{1-5}$: the southern Milky Way (the Hydra to Puppis region [D$_1$]
by Salem \& Kraan-Korteweg in prep., the Hydra/Antlia Supercluster
region [D$_2$] by Kraan-Korteweg 1999, the Crux region [D$_3$] by 
Woudt 1998, Woudt \& Kraan-Korteweg in prep., the GA region [D$_4$] by
Woudt 1998, Woudt \& Kraan-Korteweg in prep, and the Scorpius region 
[D$_5$] by Fairall \& Kraan-Korteweg in prep.); \\
E:  the Ophiuchus Supercluster by Wakamatsu \etal\ 1994, Hasegawa
\etal\ 1999; \\
F: the northern GP/SGP crossing by Hau \etal\ 1996.

These surveys have uncovered over 50\,000 previously unknown galaxies
and are not biased with respect to any particular morphological type.
Although the various optical surveys are based on different plate
material and were performed by different groups, the search techniques
overall are similar. All ZOA regions have been searched to diameter
limits of at least ${\rm D} \ga 0\farcm2$. This is considerably deeper
than the previously regarded ``whole-sky'' catalogs with their
completeness limits of ${\rm D} \ga 1\farcm3$. How can these catalogs
be merged to arrive at a well-defined whole-sky galaxy distribution
with a reduced ZOA?

\section{Completeness of Optical Galaxy Searches}

In order to merge the various deep optical ZOA surveys with existing
galaxy catalogs, Kraan-Korteweg (1999) and Woudt (1998) have analyzed
the completeness of their ZOA galaxy catalogs -- the Hy/Ant [D$_2$],
Crux [D$_3$] and GA [D$_4$] region -- as a function of the foreground
extinction.

By studying the apparent diameter distribution as a function of
the extinction E(B-V) (Schlegel \etal\ 1998) as well as the
location of the flattening in the slope  of the cumulative 
diameter curves $\log {\rm D}$-$\log{\rm N}$ for various extinction 
intervals (\cf\ Fig.~5 and 6 in Kraan-Korteweg 1999), we conclude
that our optical ZOA surveys are complete to an apparent diameter
of ${\rm D} = 14\arcsec$ -- where the diameters correspond 
to an isophote of 24.5~mag/arcsec${^2}$ (Kraan-Korteweg 1999) --
for extinction levels less than ${\rm A_B} = 3\fm0$.

How about the intrinsic diameters, \ie\ the diameters galaxies would
have if they were unobscured? A spiral galaxy seen through an
extinction of ${\rm A_B} = 1\fm0$ will, for example, be reduced to
$\sim 80\%$ of its unobscured size.  Only $\sim 22\%$ of a (spiral)
galaxy's original dimension is seen when it is observed through ${\rm
A_B} = 3\fm0$.  In 1990, Cameron derived analytical descriptions to
correct for the obscuration effects by artificially absorbing the
intensity profiles of unobscured galaxies. These corrections depend
quite strongly on morphological type due to the difference in
surface brightness profiles and mean surface brightness between
early-type galaxies and spiral galaxies. Applying these corrections, we
find that at ${\rm A_B} = 3\fm0$, an obscured spiral or elliptical
galaxy at our {\it apparent} completeness limit of ${\rm D} =
14\arcsec$ would have an intrinsic diameter of ${\rm D^o} \approx
60\arcsec$ or ${\rm D^o} \approx 50\arcsec$, respectively. At extinction
levels higher than ${\rm A_B} = 3\fm0$, an elliptical galaxy with
${\rm D^o} = 60\arcsec$ would appear smaller than the completeness
limit ${\rm D} = 14\arcsec$ and might have gone unnoticed.  The here
discussed optical galaxy catalog should therefore be complete to ${\rm
D^o} \ge 60\arcsec$ for galaxies of all morphological types down to
extinction levels of ${\rm A_B} \le 3\fm0$ with the possible exception
of extremely low-surface brightness galaxies. Only intrinsically very
large and bright galaxies -- particularly galaxies with high surface
brightness -- will be recovered in deeper extinction layers.  This
completeness limit could be confirmed by independently analyzing the
diameter vs. extinction and the cumulative diameter diagrams for
extinction-corrected diameters.

We can thus supplement the ESO, UGC and MCG catalogs -- which 
are complete to ${\rm D} = 1\farcm3$ -- with galaxies from optical 
ZOA galaxy searches that have ${\rm D^o} \ge 1\farcm3$ and ${\rm A_B}
\le 3\fm0$. As our completeness limit lies well above the ESO, UGC and 
MCG catalogs, we can assume that the other similarly performed optical
galaxy searches in the ZOA should also be complete to ${\rm D^o} =
1\farcm3$ for extinction levels of ${\rm A_B} \le 3\fm0$.

\begin{figure}
\begin{center}
\hfil \psfig{file=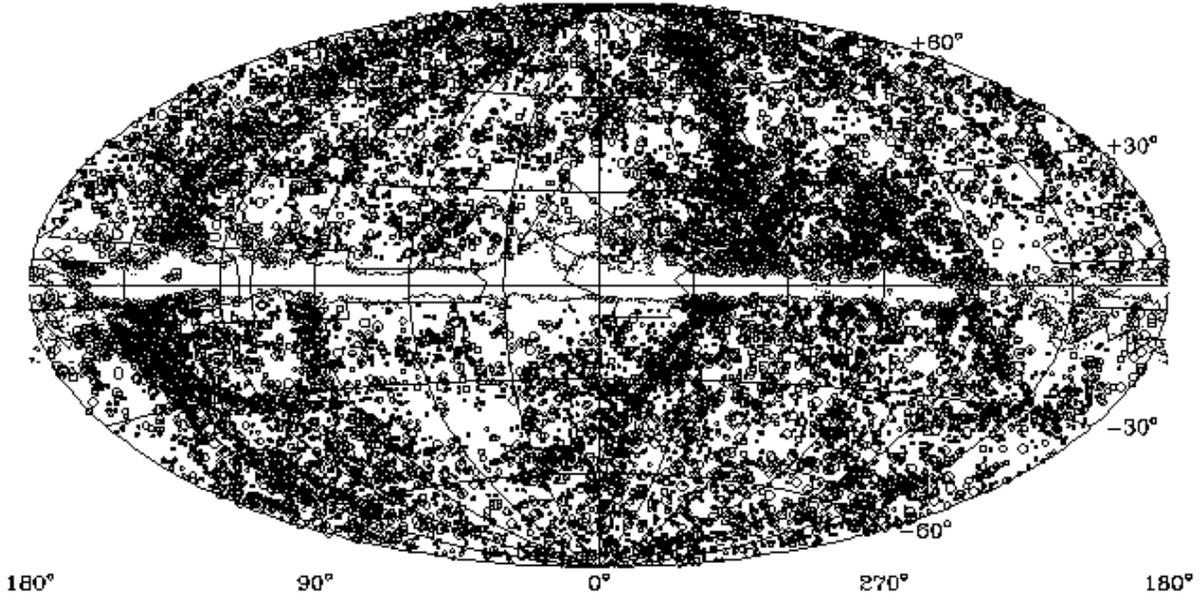,width=16cm} \hfil
\caption
{Aitoff equal-area distribution of ESO, UGC, MCG galaxies with 
extinction-corrected diameters ${\rm D^o} \ge 1\farcm3$, including 
galaxies identified in the optical ZOA galaxy searches for 
extinction-levels of ${\rm A_B} \le 3\fm0$ (contour). The diameters are
coded as in Fig.~\ref{ait}. With the exception of the areas for
which either the positions of the galaxies or their diameters are
not yet available (demarcated areas), the ZOA could be reduced
considerably compared to Fig.~\ref{ait}.}
\label{aitc}
\end{center}
\end{figure}

In Fig.~\ref{aitc} we have then taken the first step in arriving at
an improved whole-sky galaxy distribution with a reduced ZOA. In this
Aitoff projection we have plotted all the UGC, ESO, MCG galaxies that
have {\it extinction-corrected} diameters ${\rm D^o} \ge 1\farcm3$
(remember that galaxies adjacent to the optical galaxy search regions
are also affected by absorption though to a lesser extent: ${\rm
A_B} \le 1\fm0$), and added the galaxies from the various optical
surveys with ${\rm D^o} = 1\farcm3$ and ${\rm A_B} \le 3\fm0$ for
which positions and diameters were available. The regions for which
these data are not yet available are marked in Fig.~\ref{aitc}. As
some searches were performed on older generation POSS I plates,
which are less deep compared to the second generation POSS II and
ESO/SERC plates, an additional correction was applied to those
diameters, \ie\ the same correction as for the UGC galaxies which also
are based on POSS I survey material (${\rm D_{25} = 1.15 \cdot 
D_{POSS I}}$).

A comparison of Fig.~\ref{ait} with Fig.~\ref{aitc} demonstrates 
convincingly how the deep optical galaxy searches realize a 
considerable reduction of the ZOA; we can now trace the
large-scale structures in the nearby Universe to extinction levels
of ${\rm A_B} = 3\fm0$. Inspection of Fig.~\ref{aitc} reveals that
the galaxy density enhancement in the GA region is even more
pronounced and a connection of the Perseus-Pisces chain across the
Milky Way at $\ell=165\deg$ more likely. Hence, these supplemented
whole-sky maps certainly should improve our understanding of the 
velocity flow fields and the total gravitational attraction on 
the Local Group.

Redshift follow-ups of well-defined samples of ZOA galaxies
will be important in analyzing the large-scale structures in 
redshift-space. Systematic redshift surveys have been performed
for various ZOA regions and revealed a number of 
dynamically important structures such as

-- the rich, massive ($\sim 2-5 \cdot 10^{15}\msun$) cluster A3627 
at ($\ell,b,v)\sim(325\deg,-7\deg,4882~$\kms) which seems to 
constitute the previously unrecognized but predicted
density peak at the bottom of the potential well of the Great
Attractor (Kraan-Korteweg \etal\ 1996)\\
-- the 3C129 cluster at ($\ell,b,v)\sim(160\deg,0\deg,5500$~\kms) 
connecting Perseus-Pisces and A569 across
the Galactic Plane (Chamaraux \etal\ 1990, Pantoja \etal\ 1997) \\
-- and the Ophiuchus supercluster at
($\ell,b,v)\sim(0\deg,8\deg,8500$~\kms) behind the Galactic Center 
(Wakamatsu \etal\ 1994, Hasegawa \etal\ 1999).

Optical galaxy searches, however, fail in the most opaque part of the
Milky Way, the region encompassed by the ${\rm A_B} = 3\fm0$ contour
in Fig.~\ref{aitc} -- a sufficiently large region to hide further
dynamically important galaxy densities. Here, systematic \HI\ surveys
have proven very powerful as the Galaxy is transparent to the 21-cm
line radiation of neutral hydrogen and \HI-rich galaxies can readily
be found through detection of their redshifted 21-cm emission.

\section{H\thinspace{\normalsize {\bf I}} Galaxy Searches in the ZOA}

In March 1997, a systematic blind \HI\ survey began in the most opaque
part of the southern Milky Way ($213\deg \le \ell \le 33\deg$; $|b|
\le 5\fdg5$) with the Multibeam (MB) receiver (13 beams in the focal
plane array) at the 64\,m Parkes telescope. The ZOA is being surveyed
along constant Galactic latitudes in 23 contiguous fields of length
$\Delta\ell=8\deg$. The ultimate goal is 25 scans per field where
adjacent strips will be offset in latitude by $\Delta b = 1\farcm5$
for homogeneous sampling. With an effective integration time of 25
min/beam we obtain a 3\,$\sigma$ detection limit of 25\,mJy.  The
survey covers a velocity range of $-1200 \la v \la 12700$\kms\, with
a channel spacing of 13.2~\kms\ per channel, and will
be sensitive to normal spiral galaxies well beyond the Great Attractor
region.

So far, a shallow survey based on 2 out of the foreseen 25 driftscan
passages has been analyzed (\cf\ Kraan-Korteweg, Koribalski \&
Juraszek 1998, Henning
\etal\ 1999). 107 galaxies were catalogued with peak \HI-flux
densities of $\ga$80~mJy (${\rm rms} = 15$~mJy after Hanning
smoothing) and their detection show no dependence on Galactic
latitude, nor the amount of foreground obscuration through which they
have been detected.

Four cubes centered on the Great Attractor region ($300\deg \ge \ell
\ge 332\deg$, $|b| \le 5\fdg5$) of the full-sensitivity survey have
been analyzed (Juraszek \etal\ 1999) and uncovered 236 galaxies
above the $3\sigma$ detection level of 25~mJy. 70\% of the detections
had no previous identification.

\begin{figure}
\begin{center}
\hfil \psfig{file=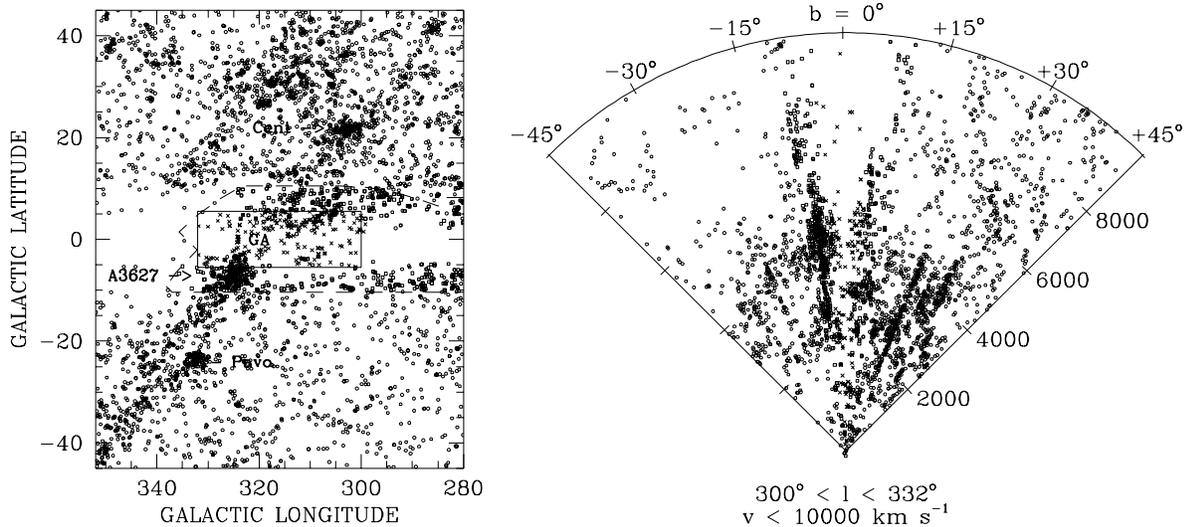,width=16cm} \hfil
\caption
{A sky distribution (left) and redshift cone (right) for galaxies with
v$<$10000~\kms in the GA region. Circles mark redshifts from the
literature (LEDA), squares redshifts from the optical galaxy search in
the Hy/Ant-Crux-GA regions (outlined on left panel) and crosses
detections in the full-sensitivity \HI\ MB-ZOA survey (box).}
\label{MBGA}
\end{center}
\end{figure}

In the left panel of Fig.~\ref{MBGA}, a sky distribution centered on
the GA region displays all galaxies with redshifts ${\rm v} \le
10000$~\kms. Next to redshifts from the literature (circles; LEDA),
redshifts from the follow-up observations of Kraan-Korteweg and
collaborators in the Hy/Ant-Crux-GA ZOA surveys (dashed area) are
plotted. They clearly reveal the prominence of the cluster A3627 at
$(\ell,b,v) = (325\deg,-7\deg,4882$~\kms, Kraan-Korteweg \etal\ 1996)
close to the core of the GA region at $(\ell,b,v) =
(320\deg,0\deg,4500$~\kms) as predicted by Kolatt \etal\ (1995).
Adding now the new detections from the systematic blind \HI\ MB-ZOA
survey (box), we for the first time can trace structures all the way
across the Milky Way. The new picture seems to suggest that the GA
overdensity is a ``great-wall'' like structure starting at the Pavo
cluster, having its core at the A3627 cluster and then bending over
towards shorter longitudes across the ZOA.

This becomes even clearer in the right panel of Fig.~\ref{MBGA} where
the galaxies are displayed in a redshift cone out to ${\rm v} \le
10000$~\kms\ for the longitude range $300\deg \le \ell \le 332\deg$
analyzed so far of the MB full-sensitivity data.  The A3627 cluster
is clearly the most massive galaxy cluster uncovered by the various
surveys in the GA region and therefore the most likely candidate for
the previously unidentified but predicted density-peak at the bottom
of the potential well of the GA overdensity.

Finding a hitherto uncharted further cluster of galaxies at the heart
of the GA would have serious implications for our current
understanding of this massive overdensity in the local Universe.
Various indications suggest that PKS1343-601, the second brightest
extragalactic radio source in the southern sky ($f_{20cm} = 79$~Jy,
McAdam 1991), might form the center of yet another highly obscured rich
cluster, particularly as it also shows significant X-ray emission (\cf\
Kraan-Korteweg \& Woudt 1999 for further details). At ($\ell, b) \sim
(310\deg,2\deg)$, this radio galaxy lies behind an obscuration layer
of about 12 magnitudes of extinction in the B-band, hence optical
surveys are ineffective.  Still, West \& Tarenghi (1989) observed this
source and identified it -- with an extinction-corrected diameter of
${\rm D^o} \approx 4\arcmin$ and a recession velocity of v = 3872~\kms\
-- as a giant elliptical galaxy. Giant ellipticals generally reside at
the cores of clusters. 

Interestingly enough, the \HI\ MB survey does
uncover a significant excess of galaxies at this position in velocity
space (\cf\ Fig.~\ref{MBGA}). However, we do not see a ``finger of
God'', the characteristic signature of a cluster in redshift space.
Could it be that too many central cluster galaxies are missed by
the \HI\ observations because spiral galaxies generally avoid
the cores of clusters? The existence of this possible
cluster still remains a mystery. Meanwhile, this prospective cluster 
has been imaged in the near infrared (Woudt \etal\ in progress), where
extinction effects are less severe compared to the optical, and which
should uncover early-type cluster members if they are there. The
forthcoming results should then unambiguously settle the question 
whether another cluster forms part of the GA supercluster.

\section*{Acknowledgements}

The collaborations with our colleagues in the optical surveys
and redshift follow-up observations, C. Balkowski, V. Cayatte, 
A.P. Fairall, C. Salem, P.A. Woudt, and the HIPASS ZOA team members
R.D. Ekers, A.J. Green, R.F. Haynes, P.A. Henning,
M. J. Kesteven, B. Koribalski, R.M. Price, E. Sadler and 
L. Staveley-Smith are greatly appreciated.

\section*{References}

\reference
Cameron L.M. 1990
A\&A 233, 16

\reference
Chamaraux P., Cayatte V., Balkowski C., Fontanelli P. 1990
A\&A 229, 340

\reference
Cardelli J.A., Clayton G.C., Mathis J.S. 1989
ApJ 345, 245

\reference
Fairall A.P., Kraan-Korteweg R.C., in prep.
[D$_5$]
%Scorpius  optical search

\reference
Hasegawa T., Wakamatsu K., Malkan M. \etal\ 1999
MNRAS, in press [E]
%optical search in Ophiuchus

\reference 
Hau G.K.T., Ferguson H.C., Lahav O. \etal\ 1996 
MNRAS 277, 125 [F]
%optical search in GP/SGP crossing

\reference
Henning P.A., Staveley-Smith L., Kraan-Korteweg R.C., Sadler E.M. 1999
PASA 16, 35

\reference
Hudson M., Lynden-Bell D. 1991
MNRAS 252, 219

\reference
Juraszek S., Staveley-Smith L., Kraan-Korteweg R.C., Green A.J., 
Ekers R.D., Henning P.A., Koribalski B.S., Sadler E.M., Schr\"oder
A.C., in prep.

\reference
Kogut A., Lineweaver C., Smoot G.F \etal\ 1993
ApJ 419, 1

\reference
Kolatt T., Dekel A., Lahav O. 1995 
MNRAS 275, 797
%center of GA at l=320, b=0 from reconstructions

\reference
Kraan-Korteweg R.C. 1999
A\&ASS, submitted

\reference
Kraan-Korteweg R.C., Woudt P.A. 1999
PASA 16, 53

\reference
Kraan-Korteweg R.C., Woudt P.A., Cayatte V. \etal\ 1996 
Nature 379, 519
%A2367 at the core of the GA

\reference
Kraan-Korteweg R.C., Koribalski B., Juraszek S. 1998
in Looking Deep in the Southern Sky, eds.
R. Morganti, W. Couch, Springer, 23

\reference
Lahav O. 1987
MNRAS 225, 213

\reference
Lauberts, A. 1982
The ESO/Uppsala Survey of the ESO (B) Atlas, ESO, Garching

\reference
Lercher G., Kerber F., Weinberger R. 1996
A\&ASS 117, 369 [B$_1$]
%POSS I R: 120-130, pm10

\reference 
Lynden-Bell D., Lahav, O. 1988
in Large-Scale Motions in the Universe, eds. V.C. Rubin and G.V. Coyne,
Princeton: Princeton University, 199

\reference
Marchiotti W., Wildauer H., Weinberger R. 1999
in progress [B$_2$]
%POSS II R : 20-80, |b|: 5-10

\reference
McAdam W.B. 1991 
PASA 9, 255

\reference
Nilson P. 1973 
Uppsala General Catalog of Galaxies,
Uppsala, University of Uppsala

\reference
Pantoja C.A., Altschuler D.R., Giovanardi C., Giovanelli R. 1997
AJ 113, 905 [A]
%optical search in PP region

\reference
Roman A.T., Nakanishi K., Tomita A., Saito M. 1996
PASJ 48, 679 [C$_3$]
%optical search in GB to l=40 region

\reference
Roman A.T., Nakanishi K., Saito M. 1998
PASJ 50, 37 [C$_2$]
%optical search in Aquila/Sagittarius region

\reference
Saito M., Ohtani A., Asomuna A. \etal\ 1990 
PASJ 42, 603 [C$_1$]
%optical search in Puppis I region (230-250)

\reference
Saito M., Ohtani A., Baba A. \etal\ 1991 
PASJ 43, 449 [C$_1$]
%optical search in Puupis II region (210-230)

\reference
Salem C., Kraan-Korteweg R.C., in prep.
[D$_1$]
%Puppis --> Hy/Ant  optical search

\reference
Saurer W., Seeberger R., Weinberger R. 1997
A\&ASS 126, 247 [B$_1$]
%POSS I R:130-180, pm5

\reference
Schlegel D.J., Finkbeiner D.P., Davis M. 1998 
ApJ 500, 525
 
\reference
Seeberger R., Saurer W., Weinberger R., Lercher G. 1994
in Unveiling Large-Scale Structures Behind the Milky Way, 
eds.\ C. Balkowski, R.C. Kraan-Korteweg, ASP Conf. Ser. 67, 81 [B$_1$]
%summary 33-240, pm5; 80-130, pm10

\reference
Seeberger R., Saurer W., Weinberger R. 1996
A\&ASS 117, 1 [B$_1$]]
%POSS I R:  180-240 pm 5

\reference
Seeberger R., Saurer W. 1998
A\&ASS 127, 101 [B$_1$]
%POSS I E: 90-100 pm10

\reference
Vorontsov-Velyaminov B., Archipova V. 1963-74
Morphological Catalog of Galaxies, Parts 2 to 5,
Moscow, Moscow University

\reference
Wakamatsu K., Hasegawa T., Karoji H. \etal\ 1994
in Unveiling Large-Scale Structures Behind the Milky Way, 
eds.\ C. Balkowski, R.C. Kraan-Korteweg, ASP Conf. Ser. 67, 131 [E]
%Ophiuchus

\reference
Weinberger R., Gajdosik M., Zanin C. 1999
A\&ASS 137, 293 [B$_5$]
%POSS II R : 113-157, |b|: 5-10

\reference
West, R.M., Tarenghi M. 1989 
A\&A 223, 61

\reference
Woudt P.A. 1998
Ph.D. thesis, Univ. of Cape Town. [D$_3$, D$_4$]
%Crux and GA search, completeness

\reference
Woudt P.A., Kraan-Korteweg R.C.,
A\&ASS, in prep. [D$_3$]
%Crux optical search

\reference
Woudt P.A., Kraan-Korteweg R.C.,
A\&ASS, in prep. [D$_4$]
%GA optical search

\end{document}